\documentclass[reprint,aps,prl,showpacs,floatfix,notitlepage,superscriptaddress]{revtex4-2}
\pdfoutput=1
\usepackage{epsfig,graphics,amssymb,amsmath,subeqnarray,color,bm,bbm}
\usepackage{graphicx}
\usepackage[toc,page]{appendix}
\usepackage{float}
\usepackage[colorlinks=true,allcolors=blue]{hyperref}

\usepackage{gensymb}

\begin{document}
\title{Escaping kinetic traps using non-reciprocal interactions}

\author{Saeed Osat}
\altaffiliation{These authors contributed equally to this work}
\email{saeed.osat@ds.mpg.de}
\affiliation{Max Planck Institute for Dynamics and Self-Organization (MPI-DS), 37077 G\"ottingen, Germany}
\author{Jakob Metson}
\altaffiliation{These authors contributed equally to this work}
\affiliation{Max Planck Institute for Dynamics and Self-Organization (MPI-DS), 37077 G\"ottingen, Germany}
\author{Mehran Kardar}
\affiliation{Department of Physics, Massachusetts Institute of Technology, Cambridge, MA 02139, United States}
\author{Ramin Golestanian}
\email{ramin.golestanian@ds.mpg.de}
\affiliation{Max Planck Institute for Dynamics and Self-Organization (MPI-DS), 37077 G\"ottingen, Germany}
\affiliation{Rudolf Peierls Centre for Theoretical Physics, University of Oxford, Oxford OX1 3PU, United Kingdom}
\date{\today}

\begin{abstract}
Kinetic traps are a notorious problem in equilibrium statistical mechanics, where temperature quenches ultimately fail to bring the system to low energy configurations. Using multifarious self-assembly as a model system, we introduce a mechanism to escape kinetic traps by utilizing non-reciprocal interactions between components. Introducing non-equilibrium effects offered by broken action-reaction symmetry in the system pushes the trajectory of the system out of arrested dynamics. The dynamics of the model is studied using tools from the physics of interfaces and defects. Our proposal can find applications in self-assembly, glassy systems, and systems with arrested dynamics to facilitate escape from local minima in rough energy landscapes.
\end{abstract}
                           
\maketitle


Since the pioneering work of H.A. Kramers \cite{Kramers1940}, studies of activated barrier crossing have found applications in a variety of situations where systems are trapped in meta-stable states, such as the stripe patterns observed in the Ising model \cite{Olejarz_PRL2012} and the coarsening dynamics of non-conserved Ginzburg-Landau equation~\cite{Bray_AP1994}. Escaping from such kinetic traps at equilibrium is typically mediated by rare nucleation events of the stable phase~\cite{Binder_RPP1987}, while exerting external stresses (at the boundaries) that drive the system away from equilibrium can also facilitate such transitions \cite{Zaccone2023}. It is known that some biological systems with intrinsic non-equilibrium activity such as enzymes are equipped with physical mechanisms that are able to effectively lower the barrier and facilitate escape \cite{Golestanian2019phoretic}. While this is an intrinsic feature for each enzyme as perfected through evolution, it is possible to shed light on how such a phenomenon could emerge through mechanochemical coupling of such molecular oscillators with stochastic barrier-crossing dynamics \cite{Agudo2021,Chatzittofi2023}. In light of this observation, it will be interesting to pose the following general question: how can we design non-equilibrium strategies for systems with many interacting degrees of freedom that can enable them to collectively overcome kinetic barriers using local free energy input? Here, we present a strategy to achieve this goal by utilizing the recently developed concept of non-reciprocal interactions in active matter as implemented to trigger dynamic shape-shifting in self-assembled structures \cite{Osat2023}.

\textit{Chimeric kinetic traps and multifarious self-assembly}.---
Kinetic traps present a significant challenge for systems trying to find energy minima.
In the context of self-assembly, kinetic traps lead to erroneous structures being formed during assembly~\cite{Whitelam_ARPC2015}. Self-assembly involves the aggregation of small building blocks to create larger structures with desired shapes and functions~\cite{Whitesides_Science2002, Glotzer_Science2004, Zhang_NL2004, Hu2007PRE, McMullen2022, Gartner2022, Gartner2024}. The design problem in self-assembly entails constructing an interaction matrix, which specifies the interactions between blocks to achieve the desired final configuration~\cite{Hormoz_PNAS2011, Nguyen2016}. However, this specificity alone does not guarantee error-free self-assembly, even at low temperatures~\cite{Whitelam_ARPC2015,Murugan2015NatComm}, as can for example be seen in the context of the so-called multifarious self-assembly (MSA) model \cite{Murugan_PNAS2015, Sartori_PNAS2020, Bisker_PNAS2018, Faran2023, Bohlin2023}, which we will use here as a model system. 

The MSA model is designed to store and retrieve multiple different structures. Here the structures, also known as patterns, correspond to particular configurations of different tile species. In MSA we are given pre-defined structures, and the goal is to design the interactions between tile species such that the system can assemble any of the given structures. This effectively turns the tile pool into an associative memory, reminiscent of a Hopfield network~\cite{Hopfield_PNAS1982, Amit_PRL1985, teixeira2023liquid}.
To successfully retrieve a structure in MSA, it is necessary to initiate the process close to the corresponding energy minimum, which can be achieved by introducing a small seed of the desired structure, or by employing concentration patterning techniques \cite{Murugan_PNAS2015, evans2024pattern}. However, since the tile pool is shared, each tile is involved in interactions with different neighboring tiles in each structure. As more structures are stored in the system, the increased cross-talk between tile species leads to the formation of many unwanted local minima corresponding to spurious chimeric structures~\cite{Amit_PRA1985, Huntley_PNAS2016}, which are analogous to the long-lived metastable states observed in spin models.

\begin{figure*}[t]
\begin{center}
\includegraphics[width=0.99\textwidth]{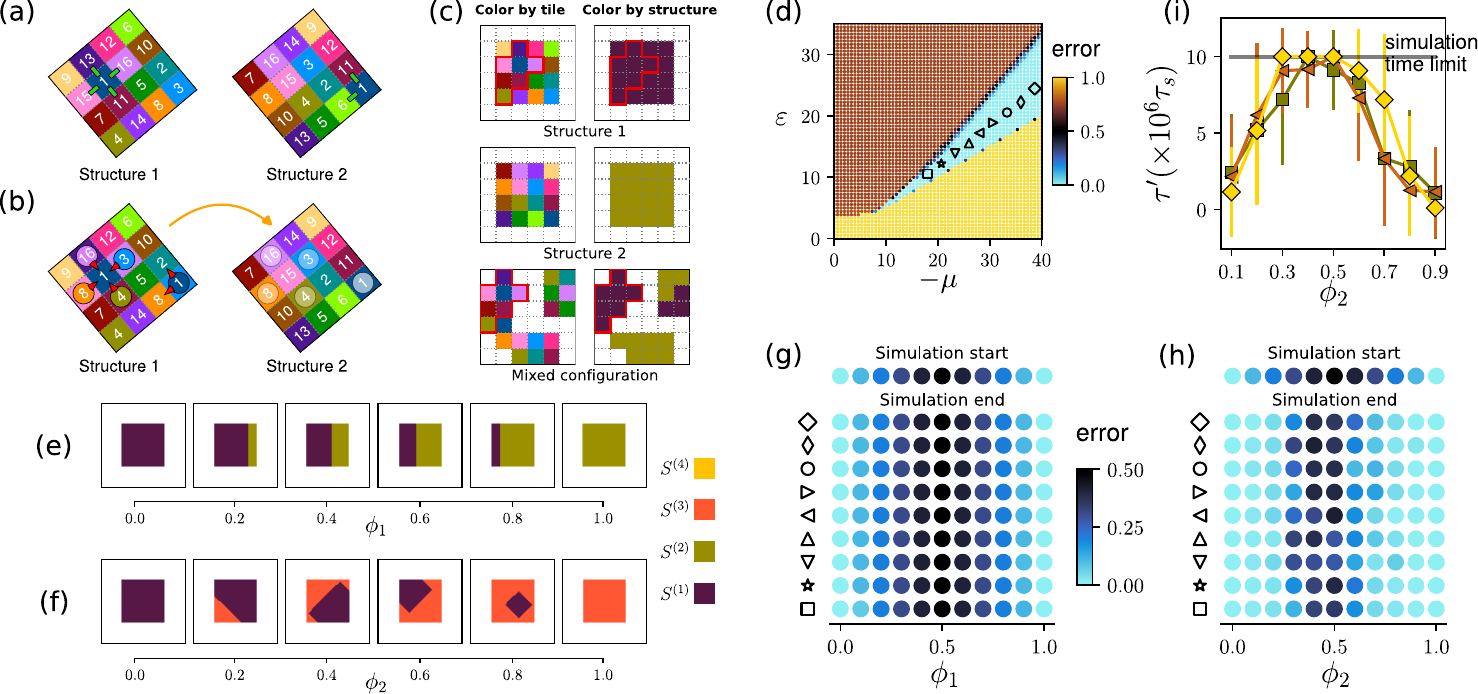}
\caption{{Non-reciprocal self-organization model and chimeric kinetic traps.} 
An illustration of all the (a) reciprocal and (b) non-reciprocal interactions involving tile species 1, in a system with two $4\times4$ structures with a shifting sequence $S^1\to S^2$. 
(c) An illustration of the coloring protocol. Completely assembled structures 1 and 2 and a mixed configuration of tiles are shown sitting in a small system. 
The red border shows a section of structure 1 colored in the mixed configuration. 
See the Supplemental Material \cite{suppmat} for details of the coloring protocol. For clarity, we generally color snapshots by structure.
(d) The $(\varepsilon, \mu)$ state diagram of the MSA model, with the cyan region indicating the regime of robust self-assembly. 
(e) Type one seeds exhibit smooth boundaries.
(f) Type two seeds have jagged boundaries.
$\phi_1$ and $\phi_2$ characterize the composition of the type one and type two chimeric seeds respectively.
(g) Average error over 10 independent realizations starting from seeds of type one.
(h) Average error starting from seeds of type two.
(i) First passage time to the nearest error-free configuration plotted against $\phi_2$.
}
\label{fig1}
\end{center}
\end{figure*}

\textit{Non-reciprocal multifarious self-organization}.---
Non-reciprocal interactions 
have found applications in many areas of active matter physics \cite{Uchida2010,Soto2014,ivlev2015Statistical,Jaime_PRL2019,Saha_PRX2020,Loos2020,You2020PNAS,Julien, Loos_PRL2023,OuazanReboul2023}. In the context of self-assembly, the non-reciprocal multifarious self-organization (NRMSO) model \cite{Osat2023} adds non-reciprocal interactions into the MSA model. This results in a unique dynamical feature where a self-assembled structure can shape-shift to the next structure in a given sequence. In this Letter, we demonstrate that by adding non-reciprocal interactions to MSA we enable the escape from metastable chimeric states that would otherwise persist for extremely long times in equilibrium conditions. 


We consider $m=4$ desired target structures, denoted $S^{\ell}$, each of which is a random permutation of $M$ distinct tiles arranged in a 2D square lattice. The simulation system is a 2D square lattice with a side lengths $[L_x,L_y]$. The structure size is chosen to be either equal to the system size or a quarter of it. The initial seed is placed at the center of the system in both cases. We use a generalized version of the grand canonical Monte Carlo method to simulate the system. For each Monte Carlo step, a random lattice point $(i, j)$ is selected, and its component $\sigma_{i, j}$ is replaced by another random component $\sigma^{\prime}$ with  probability $p={\rm min}\big\{ 1, \exp\left(\Lambda-\Delta {\cal H}\right)\big\}$ where $\Lambda = R_{\sigma_{i-1,j} \swarrow \sigma^{\prime}}^{\text{nr}}+
          R_{\sigma^{\prime} \nearrow \sigma_{i+1,j}}^{\text{nr}}+ 
          R_{\sigma^{\prime} \searrow \sigma_{i,j-1}}^{\text{nr}}+
          R_{\sigma_{i,j+1} \nwarrow \sigma^{\prime}}^{\text{nr}}$
and ${\cal H}=\sum_ {\left \langle \alpha,\beta \right\rangle } 
    U_{\sigma_{\alpha} \square \sigma_{\beta}}^{\text{r}} 
    -\mu \, n$.
$n$ is the number of tiles in the system, and all of the tiles have the same chemical potential $\mu$. $\square \in \{ \diagdown , \diagup \}$ gives the reciprocal interaction directions and $\{\swarrow, \nearrow, \searrow, \nwarrow\}$ are the non-reciprocal interaction directions (see the Supplemental Material \cite{suppmat} for more details).

The design rule, corresponding to constructing the reciprocal ($U^{\text{r}}$) and non-reciprocal ($R^{\text{nr}}$) interaction matrices, is as follows: for the reciprocal interactions, two tile species interact specifically with an interaction strength $\varepsilon$ if they are adjacent in at least one of the desired structures. The non-reciprocal interactions are chosen such that the system is able to shape-shift between structures in a predefined sequence. Take a structure in the sequence. A tile species in this structure has a non-reciprocal interaction towards tiles in neighboring positions in the previous structure in the shifting sequence. The non-reciprocal interactions have strength $\lambda$. An illustrative example of these interactions is given in Fig.~\ref{fig1}(a) and (b).
For more details of the model and interactions, see the Supplemental Material \cite{suppmat}. 

For $\frac{2}{3} \varepsilon \lesssim \lambda \lesssim \varepsilon$ robust shape-shifting behavior is observed~\cite{Osat2023}. Figure~\ref{fig1}(d) shows the error of self-assembly in the $(\varepsilon,\mu)$ parameter space with $\lambda=0$ when starting with a whole structure as the initial seed. 
However, even systems in the error-free self-assembly regime can get caught in chimeric traps, preventing assembly within practically relevant timescales.

\begin{figure}
\begin{center}
\includegraphics[scale=0.5]{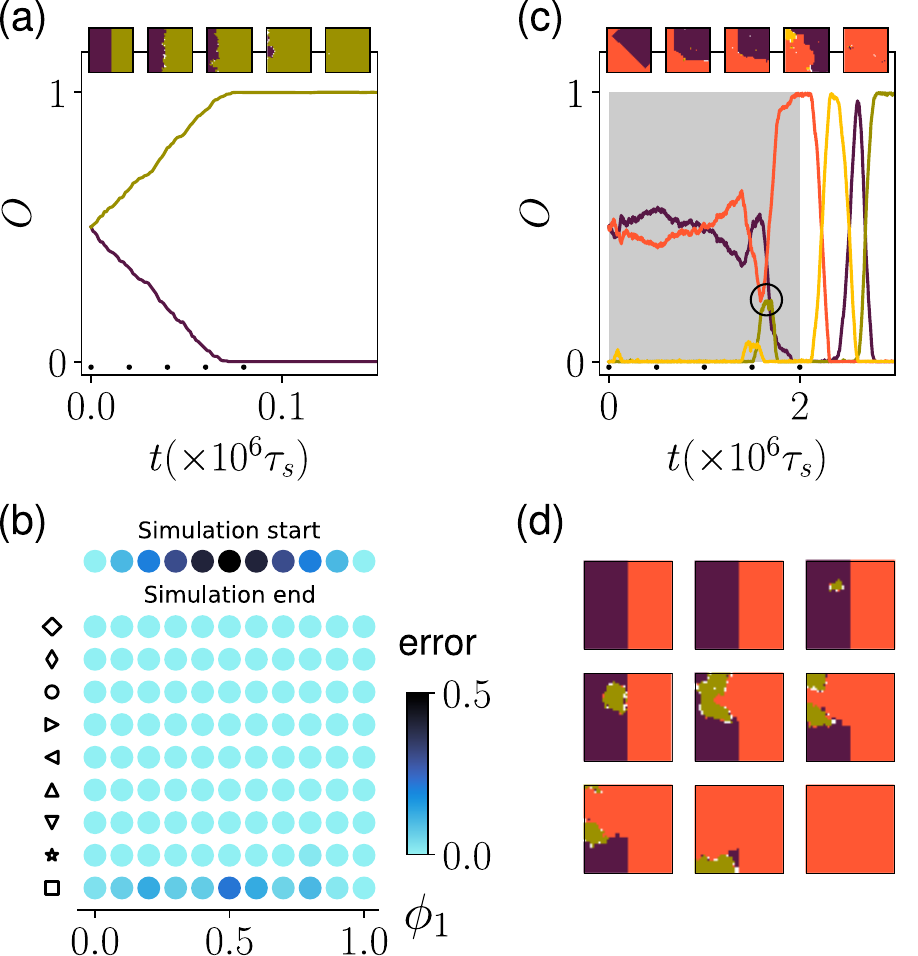}
\caption{{Escaping kinetic traps.} 
(a) A synthetic chimera seed made of two structures with a sequence $S^1 \to S^2$.
The overlap $O$ is the fraction of the system filled by each structure. Black dots on the time axis mark the times of the snapshots.
(b) Error of the final configuration for type one chimeras when non-reciprocal interactions are deployed.
(c) A chimera made of non-consecutive structures $S^1$ and $S^3$ in a cyclic sequence~$S^1 \to S^2 \to S^3 \to S^4 \to S^1$. 
(d) The same process with a non-cyclic sequence~$S^1 \to S^2 \to S^3$.}
\label{fig2}
\end{center}
\end{figure}

\textit{Escaping chimeric traps.}---
To investigate metastability in this model we begin with $\lambda =0$ (reciprocal interactions only) and define two different types of chimeric seed. Two structures are separated by a boundary that can be either commensurate with the lattice structure (type one chimera) or not (type two chimera), as shown in Fig.~\ref{fig1}(e) and (f). Initially, a fraction $\phi_i$ of the seed is assigned to one structure and a fraction $1-\phi_i$ to another structure. Here $i=\{1,2\}$ corresponds to the two different chimera types. We probe the behavior of the system starting from these seeds and choosing the values of $(\varepsilon,\mu)$ within the error-free region of the MSA model. If the dynamics of the system changes the initial seed to a completely assembled single structure, then it is considered an escape from the kinetic trap; otherwise, the dynamics is arrested in the chimeric state. Figure~\ref{fig1}(g) shows the error of self-assembly, starting from type one chimeras as the initial seed, for different $(\varepsilon, \mu)$ coordinates as marked in panel (d). As expected, smooth boundaries are robust, and the system cannot escape from this type of chimera, even for values of $\phi_1$ close to 0 or 1. On the other hand, Fig.~\ref{fig1}(h) shows the same simulations, but where the system is initialized with type two chimeras as the initial seed. The rugged interface shape facilitates the dynamics of chimeras and for both small and large $\phi_2$, the trajectory of the system is absorbed into the closest minimum (fully assembled target structure). However, for medium values of $\phi_2$, the dynamics is usually trapped in an alternate chimeric state (despite the smoothing of the rugged interfaces) or occasionally in blinker states~\cite{Olejarz_JSM2013}.
Note that MSA not only stores different structures as energy minima, but it also provides basins around these minima.
Chimeras close to the original structures can reach the original structures, and require less time to do so the closer they are. This can be seen 
in Fig.~\ref{fig1}(i).

We now demonstrate how tiles with non-reciprocal interactions can escape from the different chimeric traps introduced in Figs.~\ref{fig1}(e) and (f). 
We observe that for a sequence $S^1 \to S^2$ a chimeric seed will start to expand the $S^2$ region at the expense of diminishing the $S^1$ region, as shown in Fig.~\ref{fig2}(a). Figure~\ref{fig2}(b) presents the same simulations as Fig.~\ref{fig1}(g), but with tiles also exhibiting non-reciprocal interactions.

Through extensive simulations, we have observed two different mechanisms by which NRMSO drives the dynamics of the system out of an arrested state. The escape involves either the roughening of a domain wall, as shown in the previous example, or a combination of roughening and the dynamics of point defects as shown in Fig.~\ref{fig2}(c) and (d).
We investigate these two mechanisms in the following sections (see \cite{suppmat} for more details).

\begin{figure}
\begin{center}
\includegraphics[scale=0.65]{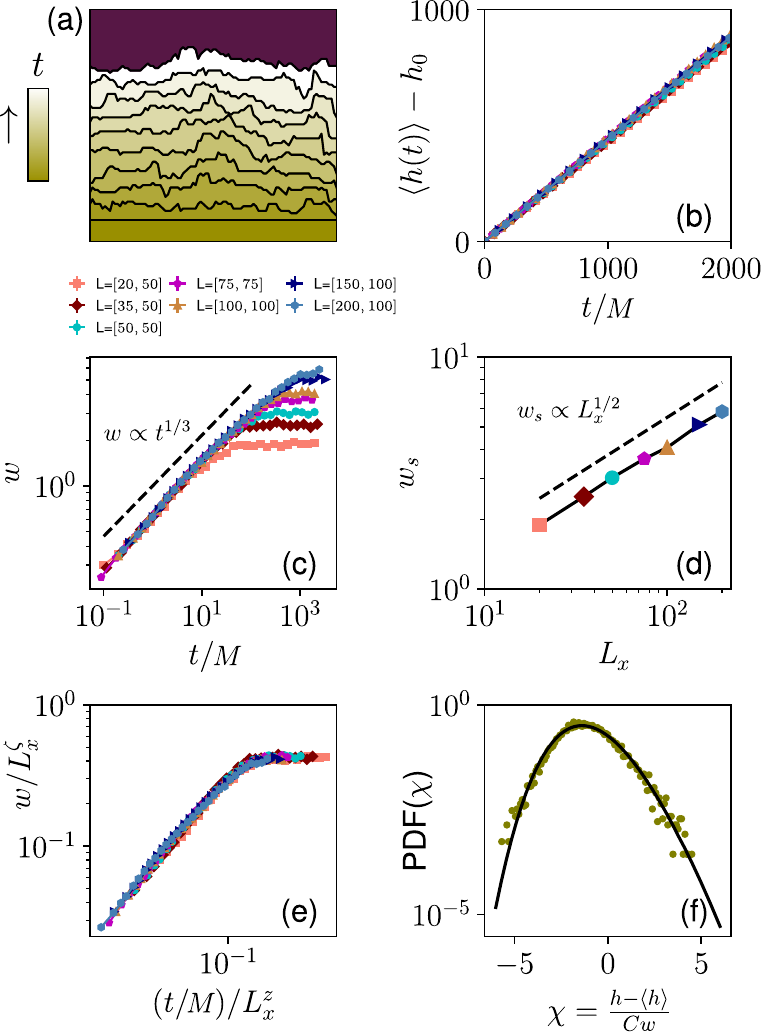}
\caption{{Interface roughening in chimeras.} 
(a) Dynamics of a chimera formed of $S^1$ and $S^2$ with a shifting sequence $S^1 \to S^2$ and an initially smooth boundary.
(b) Average interface height for different system sizes.
(c) Interface roughness as a function of time. 
(d) Saturation roughness as a function of system size. 
(e) Scaling and collapse of the data for different system sizes with roughness exponent $\zeta=1/2$ and dynamic exponent $z=3/2$.
(f) Density distribution of re-scaled height fluctuations $\chi = (h-\langle h \rangle)/(C w)$ (where $C$ is a fitting parameter; here $C=0.9$) and the corresponding GOE Tracy-Widom distribution
for a system size of $100\times 100$ in the saturation regime.}
\label{fig3}
\end{center}
\end{figure}

\textit{Interface roughening.}---
Two structures are stored, with a shifting sequence $S^1 \to S^2$. Starting from a type one chimera, which consists of $S^1$ and $S^2$ separated by a smooth boundary, and choosing a suitable $\lambda$ value, the interface begins to grow, as depicted in Fig.~\ref{fig3}(a). Since the interface moves at a constant velocity, with a finite system size we quickly reach a system filled with $S^2$. Since there can be no interface without any $S^1$ present, interface roughening and roughness saturation are no longer observed. 

We can study the dynamics of this moving interface by employing a two-phase method which overcomes the limitation of finite system size (see the Supplemental Material \cite{suppmat} for details and Supplemental Video 1) \cite{Devillard_EPL1992}. 
Simulations on a rectangular lattice with $L_y \gg L_x$ confirm the same results obtained using the two-phase method. Supplemental Video 2 shows such a simulation. Figure~\ref{fig3}(b) demonstrates that the interface moves with a constant velocity, and that this velocity remains constant for all system sizes (see also Supplemental Video 3.) Note that $t$ is rescaled by the number of tile species. 
Figure~\ref{fig3}(c) displays the roughness of the interface, defined as $w(t)=\sqrt{\langle(h(x,t) - \langle h(x,t)\rangle_x)^2 \rangle_x}$, for different system sizes. All the curves exhibit an initial growth exponent of $\beta=1/3$, followed by a crossover to a saturation regime with a size-dependent value saturation roughness. The saturated roughness increases with system size as $w_s \propto L_x^\zeta$, with a roughness exponent of $\zeta=1/2$, as shown in Fig.~\ref{fig2}(d). These exponents suggest a Kardar-Parisi-Zhang (KPZ) universality (corresponding with a dynamic exponent $z=3/2$)~\cite{KPZ, barabasi_stanley_1995,pisegna2024emergent}. Figure~\ref{fig3}(e) further confirms the KPZ universality through scaling and collapse of the curves for different system sizes. Additionally, the probability density of height fluctuations follows the Gaussian orthogonal ensemble (GOE) Tracy-Widom distribution \cite{Tracy1996} as shown in Fig.~\ref{fig3}(f), as expected from the KPZ universality class~\cite{CORWIN_2012, Takeuchi_PRL2010}.

\begin{figure}
\begin{center}
\includegraphics[scale=0.65]{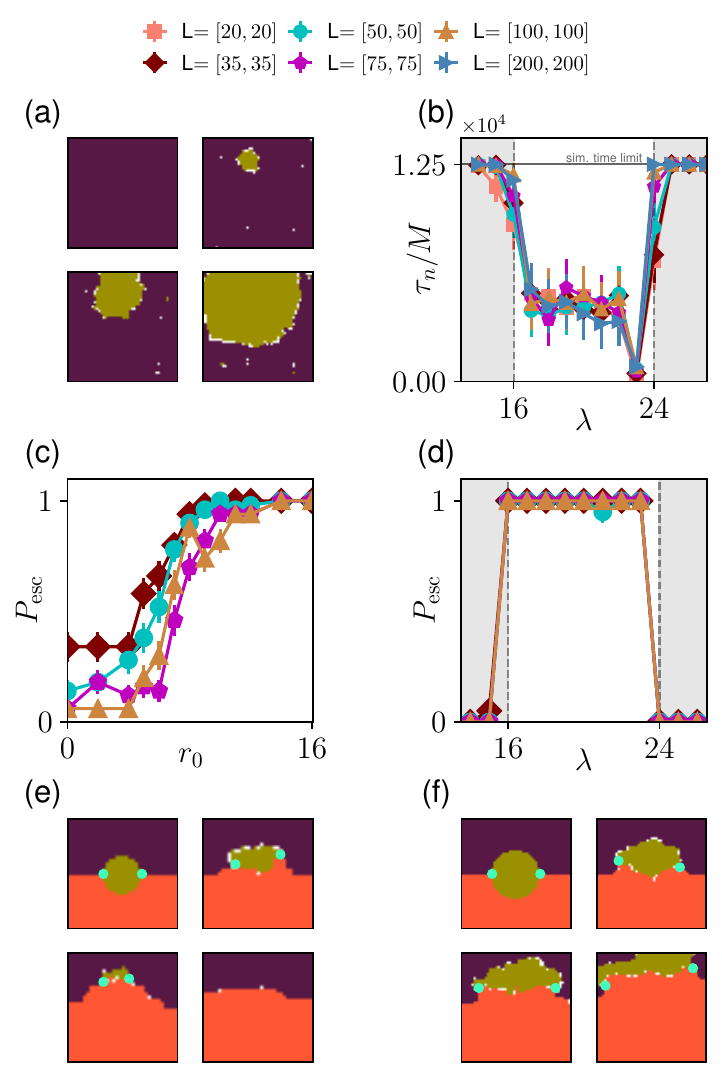}
\caption{{Defect-mediated escape.} (a) Nucleation of $S^2$ in a sea of $S^1$, with sequence $S^1 \to S^2$.  (b) Nucleation time of this process.
Probability of escape from the droplet configuration as a function of (c) initial droplet radius $r_0$ and (d) $\lambda$. 
(e) Unsuccessful and (f) successful escape from the droplet configuration. Light blue dots mark defects where the regions of $S^1,S^2,S^3$ meet. Videos of (a), (e), and (f) are included as Supplemental Videos 4, 5, and 6, respectively.}
\label{fig4}
\end{center}
\end{figure}

\textit{Defect nucleation.}---
The second mechanism of escape is via defect nucleation. This occurs when the system is trapped in a configuration where there is no direct route for the interface to grow. With a sequence like $S^1 \to S^2 \to S^3$, the escape from a seed composed of $S^1$ and $S^3$ can proceed via an initial nucleation of a droplet of $S^2$ within the $S^1$ region. The droplet grows until it reaches the boundary, upon which defects are formed. This process is demonstrated in \ref{fig2}(d). We note that defects have been shown to play an important role in the dynamics of other non-reciprocal systems \cite{rana2023defect}.

We measure the nucleation time $\tau_n$ by starting from a system filled with $S^1$ and measuring the first passage time of reaching $O_2=0.25$, where $O_2$ is the overlap of the system configuration with structure $S^2$. The system has a shifting sequence $S^1 \to S^2$. An example of a nucleation event is shown in Fig.~\ref{fig4}(a) and Supplemental Video 4. Figure~\ref{fig4}(b) shows $\tau_n$ rescaled by $M$ for different system sizes, confirming previous results on the functional region $\frac{2}{3}\varepsilon \lesssim \lambda \lesssim \varepsilon$ where shifting occurs~\cite{Osat2023}. 

Having demonstrated finite nucleation times for appropriate values $\lambda$, we now consider systems initialised with defects.
In the droplet configuration, the system is initialised with an $S^2$ droplet sitting on an interface between $S^1$ and $S^3$, with shifting sequence $S^1 \to S^2 \to S^3$, as shown in the top left panels of Figs.~\ref{fig4}(e) and \ref{fig4}(f). 
We measure the probability of escape $P_\text{esc}$, defined as the probability of reaching $O_3=0.95$ by the end of a long simulation, for different initial droplet radii $r_0$ and different values of $\lambda$. As shown in Fig.~\ref{fig4}(c) we observe a critical initial droplet radius above which the system will almost always successfully escape. Droplets smaller than this radius can be destroyed by the chasing $S^3$ structure, leading to the disappearance of the point defects, which leaves the system trapped in a chimeric state, as shown in Fig.~\ref{fig4}(e) and Supplemental Video 5. For larger droplets, the point defects spiral apart and drive the system out of the chimeric trap, which can be seen in Fig.~\ref{fig4}(f) and Supplemental Video 6.
For escape from the droplet configuration we again find the function range $\frac{2}{3}\varepsilon \lesssim \lambda \lesssim \varepsilon$, as shown in Fig.~\ref{fig4}(d). 

\textit{Conclusion.}--- We introduced an internally driven method to automatically escape kinetic traps in self-assembly using non-reciprocal interactions between the building blocks. The dynamics of the escape is quantified using tools from the physics of interfaces and defects. 
Our findings are general and are applicable to a diverse range of systems with arrested dynamics such as glassy systems~\cite{Kirkpatrick_PRB1987,Parisi_2020}, where we expect non-reciprocal interactions to accelerate the dynamics by introducing new pathways, or help to eliminate the glassy state altogether \cite{Crisanti1988PRA}. 

\acknowledgements
We acknowledge support from the Max Planck School Matter to Life and the MaxSynBio Consortium which are jointly funded by the Federal Ministry of Education and Research (BMBF) of Germany and the Max Planck Society.

S.O. and J.M. contributed equally to this work.

\bibliography{refs_main}

\end{document}